\documentclass[twocolumn,twocolappendix]{aastex631}

\newcommand{\Msolar}{M${_\odot}$}

\received{December 1, 2023}
\accepted{September 4, 2024}

\begin{document}

\title{Protoplanetary discs around sun-like stars appear to live longer when the metallicity is low\footnote{Based on observations made with the NASA/ESA/CSA James Webb Space Telescope}}

\correspondingauthor{Guido De Marchi} 
\email{gdemarchi@esa.int}

\author[0000-0001-7906-3829]{Guido De Marchi}
\affiliation{European Space Research and Technology Centre, Keplerlaan 1, 2200 AG Noordwijk, Netherlands}
\author[0000-0002-9262-7155]{Giovanna Giardino}
\affiliation{ATG Europe for the European Space Agency, European Space Research and Technology Centre, Noordwijk, Netherlands}
\author[0000-0002-1892-2180]{Katia Biazzo}
\affiliation{INAF Osservatorio Astronomico di Roma, Via Frascati 33, 00078 Monteporzio Catone, Rome, Italy}
\author[0000-0002-9309-9737]{Nino Panagia}
\affiliation{Space Telescope Science Institute, 3700 San Martin Drive, Baltimore, MD 21218, USA}
\author[0000-0003-2954-7643]{Elena Sabbi}
\affiliation{Gemini Observatory/NSF’s NOIRLab, 950 North Cherry Avenue, Tucson, AZ, 85719, US}
\affiliation{Space Telescope Science Institute, 3700 San Martin Drive, Baltimore, MD 21218, USA}
\author[0000-0002-6881-0574]{Tracy L. Beck}
\affiliation{Space Telescope Science Institute, 3700 San Martin Drive, Baltimore, MD 21218, USA}
\author[0000-0002-9573-3199]{Massimo Robberto}
\affiliation{Space Telescope Science Institute, 3700 San Martin Drive, Baltimore, MD 21218, USA}
\author[0000-0002-6091-7924]{Peter Zeidler}
\affiliation{AURA for the European Space Agency, ESA Office, STScI, 3700 San Martin Drive, Baltimore, MD 21218, USA}
\author[0000-0003-4870-5547]{Olivia C. Jones}
\affiliation{UK Astronomy Technology Centre, Royal Observatory, Blackford Hill, Edinburgh, EH9 3HJ, UK}
\author[0000-0002-0522-3743]{Margaret Meixner}
\affiliation{Jet Propulsion Laboratory, California Institute of Technology, 4800 Oak Grove Drive, Pasadena, CA 91109, USA}
\affiliation{SOFIA Science Center, USRA, NASA Ames Research Center, M.S. N232-12, Moffett Field, CA 94035, USA}
\author[0000-0003-2902-8608]{Katja Fahrion}
\affiliation{European Space Research and Technology Centre, Keplerlaan 1, 2200 AG Noordwijk, Netherlands}
\author[0000-0002-2667-1676]{Nolan Habel}
\affiliation{Jet Propulsion Laboratory, California Institute of Technology, 4800 Oak Grove Drive, Pasadena, CA 91109, USA}
\affiliation{SOFIA Science Center, USRA, NASA Ames Research Center, M.S. N232-12, Moffett Field, CA 94035, USA}
\author[0000-0002-7512-1662]{Conor Nally}
\affiliation{Institute for Astronomy, University of Edinburgh, Blackford Hill, Edinburgh, EH9 3HJ, UK}
\author[0000-0002-2954-8622]{Alec S. Hirschauer}
\affiliation{Space Telescope Science Institute, 3700 San Martin Drive, Baltimore, MD 21218, USA}
\author[0000-0002-0322-8161]{David R. Soderblom}
\affiliation{Space Telescope Science Institute, 3700 San Martin Drive, Baltimore, MD 21218, USA}
\author[0000-0001-6576-6339]{Omnarayani Nayak}
\affiliation{Goddard Space Flight Center, 8800 Greenbelt Road, Greenbelt, MD 20771, USA}
\author[0000-0003-4023-8657]{Laura Lenki\'{c}}
\affiliation{Jet Propulsion Laboratory, California Institute of Technology, 4800 Oak Grove Drive, Pasadena, CA 91109, USA}
\affiliation{SOFIA Science Center, USRA, NASA Ames Research Center, M.S. N232-12, Moffett Field, CA 94035, USA}
\author[0000-0001-5742-2261]{Ciaran Rogers}
\affiliation{Leiden Observatory, Leiden University, Leiden, the Netherlands}
\author[0000-0001-9737-169X]{Bernhard Brandl}
\affiliation{Leiden Observatory, Leiden University, Leiden, the Netherlands}
\author[0000-0002-4834-369X]{Charles D. Keyes}
\affiliation{Space Telescope Science Institute, 3700 San Martin Drive, Baltimore, MD 21218, USA}

\begin{abstract}

Previous Hubble Space Telescope (HST) observations of the star-forming cluster NGC\,346 in the Small Magellanic Cloud (SMC) had revealed a large population of pre-main--sequence (PMS) candidates, characterised by H$\alpha$ excess emission in their photometry. However, without access to spectroscopy, the nature of these objects remained unclear. Using the NIRSpec instrument on board JWST, we studied a sample of these stars, with masses in the range $\sim0.9-1.8$\,\Msolar, effective temperatures ($T_{\rm eff}$) in the range  $4,500-8,000$\,K, and PMS ages between $\sim0.1$ and 30\,Myr. Here, we present the first spectra of solar-mass PMS stars in the metal-poor SMC ($Z=1/8\, Z_\odot$) and discuss the physical properties of ten representative sources with good signal-to-noise ratio. The observations {indicate} that even the oldest of these PMS candidates are still accreting gas with typical rates of $\sim10^{-8}\,M_\odot$\,yr$^{-1}$ for stars older than $\sim10$\,Myr, {confirming their PMS nature}. The spectra also reveal near-infrared excess and molecular hydrogen excitation lines {consistent with} the presence of discs around these stars. These findings {suggest that in a low-metallicity environment circumstellar discs can live longer than previously thought}.

\end{abstract}



\section{Introduction} 
\label{sec:introduction}

In their final formation stages, after dissipating most of the envelope of dust and gas in which they were born, sun-like stars still have a mostly gaseous disc from which they continue to accrete while the star contracts until hydrogen ignites in the core  \citep{hayashi1966, shu1987, palla1999, larson2003}. It is in these discs that planetary systems form \citep{armitage2018}. In nearby Galactic star-forming regions with solar composition, the discs are short-lived, with decay timescales of $\sim 2$\,Myr and an accretion phase that rapidly tapers off \citep{haisch2001,hernandez2008,mamajek2009,fedele2010,richert2018}.

{Since the majority of stars in the Universe formed at cosmic noon around redshift $z \sim 2$ \citep{madau1996,lilly1996}, it is important to investigate the lifetime of protoplanetary discs in the lower-metallicity conditions ($Z \sim 0.1\,Z_\odot$) characteristic of those environments (e.g., \citealt{zwaan2005,fynbo2006}). In this respect, it has been suggested that lifetimes in lower-metallicity environments should be even shorter than in the solar neighbourhoods because discs are less shielded by dust from the X-ray emission of the central star \citep{ercolano2008,ercolano2009}. The fraction of disc-bearing stars with near-infrared (NIR) excess measured in two small low-metallicity ($\sim0.1\,Z_\odot$) star-forming regions in the extreme outer Galaxy initially \citep{yasui2009,yasui2010} appeared to support this hypothesis (but see Section\,6).}

In contrast, photometric studies of equally low-metallicity star-forming regions in the Magellanic Clouds using the { Hubble Space Telescope} ({ HST}) have revealed thousands of stars that appear to systematically accrete at higher rates and for times exceeding 30~Myr \citep{demarchi2010,demarchi2011a,demarchi2013,demarchi2017,biazzo2019,tsilia2023}. In particular, observations \citep{sabbi2007} of the massive star-forming region NGC\,346, in the Small Magellanic Cloud (SMC), revealed approximately $\sim690$ objects with strong H$\alpha$ excess emission and H$\alpha$ equivalent widths (EW) larger than 20\,\AA\ \citep{demarchi2011a}. As such, these objects are candidate pre-main-sequence (PMS) stars actively accreting matter from their circumstellar discs. {\cite{demarchi2011a} find that these PMS objects form two distinct populations, with a clearly bimodal age distribution: about half of the objects are younger than $\sim 2$\,Myr and the other half older than $\sim8$\,Myr, including over 200 objects with ages in the range $15-30$\,Myr {(see Section 5 for details)}. }
%There are two distinct groups of PMS stars, occupying different positions both in the colour-magnitude diagram (CMD) and on the sky, corresponding to two distinct populations with a clearly bimodal age distribution: about half of the objects are younger than $\sim 2$\,Myr (with  effective temperature $T_{\rm eff}<4,500$\,K) and the other half older than $\sim8$\,Myr (with $T_{\rm eff}>7,500$\,K), including over 200 objects with ages in the range $15-30$\,Myr {(see Section 5 for details)}. 

Since in Galactic star-forming regions accreting PMS stars older than 15\,Myr are very rare \citep{herczeg2014}, it has been proposed \citep{demarchi2010, demarchi2011a, demarchi2013, demarchi2017} that {the PMS accretion phase lasts longer} in the low-metallicity environments typical for the Magellanic Clouds ($\sim0.1-0.3\,Z_\odot$), and by inference at cosmic noon. {{While other processes such as chromospheric activity or binary interactions might be responsible for the H$\alpha$ excess emission in { HST} photometry of stars hotter than $\sim 7,500$\,K, this is unlikely because the measured H$\alpha$ emission EW is more than an order of magnitude higher than for chromospherically active binaries \citep{montes1995}}}. However, without a means to obtain spectra for these objects, the {nature of the H$\alpha$ excess emission} {of the sources in NGC\,346} has so far remained unclear.

With the NIRSpec Micro Shutter Array (MSA) we studied a sample of these candidate PMS objects in NGC\,346, as well as normal main sequence (MS) stars with no detectable H$\alpha$ excess emission. Here, we present the spectra and discuss the physical properties of ten representative sources with similar masses ($\sim0.9-1.8$\,\Msolar), $T_{\rm eff}$ in the range $4,500–8,000$\,K, and spanning a range of ages ($\sim0.1-30$\,Myr).

The structure of this paper is as follows. The observations are presented in Section\,2, while the extraction and calibration of the spectra are discussed in Section\,3. The spectral analysis is contained in Section\,4, where we derive the physical parameters of the objects. {A detailed discussion of the age of PMS stars and their uncertainties is provided in Section\,5.} Conclusions are presented in Section\,6.

\section{Observations and sample selection}
\label{sec:observations}

The central regions of NGC\,346 were observed with NIRSpec on 2022 July 26 using the G235M/F170LP medium-resolution grating ($R\sim 1000$) covering the range $1.7-3.2$\,$\mu$m (proposal number 1227). A total of 40 sources were observed simultaneously through the NIRSpec MSA, of which half had been identified as candidate PMS stars \citep{demarchi2011a}, owing to their $H\alpha$ excess emission, and the other half as candidate MS stars (no detectable excess in $H\alpha$). The targets were selected from a catalogue of about 80,000 sources in this field detected and measured with {HST} photometry \citep{sabbi2007}, out of which about 690 have $H\alpha$ excess emission exceeding $5\,\sigma$ \citep{demarchi2011a}.

Using the eMPT software tool \citep{bonaventura2023}, we identified the most suitable targets (and corresponding MSA configuration) for the specific orientation of the telescope at the time of the observations. {The MSA contains about 250,000 microshutters, each one defining a slit with a width of $0\farcs20$ and a length of $0\farcs46$, while the pitch between microshutters is $0\farcs27$ in the spectral direction and $0\farcs53$ in the spatial direction \citep{ferruit2022}.} We elected to reserve for each source a ``slitlet'' composed of three neighbouring microshutters in the cross-dispersion direction, in order to also properly characterise the nebular background around our targets. Each slitlet spans a region of $0\farcs2 \times 1\farcs5$ on the sky. Several constraints were applied to derive the final list of 40 targets, one of which was to avoid sources with contaminants within three microshutters for any position of the source in the slitlet. {Indicating with $m_I$ the $I$ band magnitude of the possible target source in the catalogue by \cite{sabbi2007}, we considered as viable targets only those for which any contaminants within the three microshutters were fainter than $m_I+5$.} We further constrained the eMPT tool to exclude objects for which the wavelengths corresponding to the Pa\,$\alpha$ or Br\,$\beta$ spectral lines would fall in the gap between detectors to ensure that both lines can be measured simultaneously. For some sources, the wavelength range around the Br\,$\gamma$ line is also recorded. 

Each source was observed at three different nodding positions in the slitlet by slewing the telescope in the cross-dispersion direction by an amount corresponding to the microshutter pitch ($0\farcs53$). Each target star was observed for 481~s in each of the three microshutters, at the same relative position, for a total 1443~s of integration. Due to the fixed MSA pattern, it is not possible to locate all targets at the nominal centre of a microshutter. 

\begin{deluxetable*}{lllccccccccccccc}[h]
\tabcolsep=0.05cm
\footnotesize
\tablecaption{{Properties of the stars in our sample}.\label{tab1} }
\tablehead{
\colhead{ID} & \colhead{RA} & \colhead{DEC} & \colhead{Age} & \colhead{$M_*$} & \colhead{$A_V$} & \colhead{$\log \,T$} & \colhead{$\log \,L$} & \colhead{$R_*$} & \colhead{$m_{200}$} & \colhead{$W_{\rm Pa \alpha}$} & \colhead{$W_{\rm Br \beta}$} & \colhead{$W_{\rm Br \gamma}$} & \colhead{$\log \,L_{\rm a}^{\rm H \alpha}$} &\colhead{$\log \,L_{\rm a}^{\rm Br \gamma}$} &  \colhead{$\log \,\dot M$}\\
   &    &     & Myr & $M_\odot$ & mag & $K$ & $L_\odot$ & $R_\odot$ & mag & \AA\, & \AA\, &
\AA\, & $L_\odot$ & $L_\odot$ & $M_\odot$\,yr$^{-1}$
}
\decimalcolnumbers
\startdata
9884  &$ 14.8247272$&$ -72.1544850$&$  0.1$&$ 0.9$&$ 0.55$&$ 3.65$&$
1.10$&$ 5.8$&$20.4$&$ -142.1$&$  -41.8$&$ -21.5$&$  1.02$&$  1.28$&$ -5.30$ \\
9685  &$ 14.8206427$&$ -72.1862390$&$  1.5$&$ 1.8$&$ 0.60$&$ 3.74$&$
0.96$&$ 3.3$&$21.6$&$  -65.1$&$  -27.1$&$ -10.6$&$  0.25$&$  0.39$&$ -6.74$ \\
24897 &$ 14.8067222$&$ -72.1913565$&$  5.5$&$ 1.2$&$ 0.65$&$ 3.75$&$
0.42$&$ 1.7$&$23.2$&$ -210.0$&$ -302.0$&$ -25.8$&$ -0.09$&$ -0.27$&$ -7.51$ \\
12559 &$ 14.8044050$&$ -72.1929702$&$ 12.5$&$ 1.3$&$ 0.66$&$ 3.91$&$
0.76$&$ 1.3$&$23.3$&$ -157.8$&$ -389.3$&$  -8.0$&$ -0.27$&$ -0.99$&$ -8.38$ \\
20408 &$ 14.7178504$&$ -72.1690186$&$ 16.0$&$ 1.2$&$ 0.56$&$ 3.89$&$ 
0.47$&$ 1.0$&$24.0$&$ -165.2$&$ -139.6$&$      $&$ -0.32$&       &          \\ 
26651 &$ 14.8120579$&$ -72.1876153$&$ 20.0$&$ 1.0$&$ 0.60$&$ 3.82$&$
0.31$&$ 1.1$&$23.9$&$ -195.0$&$ -129.7$&$  -7.6$&$ -0.59$&$ -1.41$&$ -8.76$ \\
40571 &$ 14.8088971$&$ -72.1823255$&$ 27.0$&$ 0.9$&$ 0.55$&$ 3.80$&$
0.03$&$ 0.8$&$24.4$&$ -387.4$&$ -290.0$&$ -45.4$&$ -0.10$&$ -0.41$&$ -7.68$ \\
10147 &$ 14.8854390$&$ -72.1667293$&$	  $&$ 1.3$&$ 0.56$&$ 3.89$&$ 
0.86$&$ 1.5$&$23.1$&         &         &          &      &       &          \\
14090 &$ 14.8811044$&$ -72.1616790$&$     $&$ 1.4$&$ 0.55$&$ 3.88$&$
0.68$&$ 1.3$&$23.4$&         &         &          &      &       &          \\
19639 &$ 14.8660216$&$ -72.1617760$&$     $&$ 1.2$&$ 0.42$&$ 3.84$&$
0.45$&$ 1.2$&$23.9$&         &         &          &      &       &          \\
% 11662 &$ 14.8712565$&$ -72.1820824$&$	  $&$ 1.3$&$ 0.57$&$ 3.90$&$ 
% 0.79$&$ 1.3$&&         &         &          &      &       &          \\
% 13631 &$ 14.8802135$&$ -72.1656907$&$	  $&$ 1.2$&$ 0.49$&$ 3.85$&$ 
% 0.68$&$ 1.5$&&         &         &          &      &       &          \\
\enddata
\tablecomments{The columns correspond to: source identification (ID), J2000.0 coordinates (RA, DEC), age, stellar mass ($M_*$), extinction ($A_V$), effective temperature ($\log \,T$), bolometric luminosity ($\log \,L$), stellar radius ($R_*$), magnitude in the F200W NIRCam band ($m_{200}$), EW of the Pa\,$\alpha$, Br\,$\beta$, and Br\,$\gamma$ lines, accretion luminosity derived from the $L(H\,\alpha)$ and $L(Br\,\gamma)$ luminosity, and mass accretion rate ($\log \,\dot M$). Stellar parameters are from \cite{demarchi2011a}, NIRCam F200W AB magnitudes are from \cite{habel2024} (F277W AB magnitudes for stars 9685 and 40571). All spectral lines have negative EW, indicating emission. }
\end{deluxetable*}

In this first work we present the results for ten representative sources with similar masses ($0.9 - 1.8$\,$M_\odot$, median $1.2$\,$M_\odot$) that span a range of ages, including {both younger and older} PMS candidate stars, as well as MS objects. The stars were selected to have  signal-to-noise ratio (SNR) $> 3$ (as measured {per pixel} with the {\em DER\_SNR} routine; \citealt{stoehr2008}), to be free from contaminants in the immediate vicinity of the selected microshutters (some of the contaminants were not detectable in the optical { HST} images but are visible in the NIRCam exposures; \citealt{jones2023}, \citealt{habel2024}), and for which also the Br\,$\gamma$ is recorded in the spectrum (only for one of the objects is this not the case). In Table\,\ref{tab1} we list the coordinates and key stellar parameters. The masses and ages of the PMS candidates were determined by \cite{demarchi2011a} by comparing the effective temperature $T_{\rm eff}$ and bolometric luminosity $L_{\rm bol}$ of the stars, derived from extinction-corrected $V$ and $I$ broadband {HST} photometry, to theoretical PMS evolutionary tracks \citep{deglinnocenti2008,tognelli2011} for metallicity $Z=1/8\, Z_\odot$, as appropriate for NGC\,346 (for more details, see \citealt{demarchi2011a} {and Section\,5}). For candidate MS stars, which were selected among objects with no $H\alpha$ excess emission in the catalogue of \cite{demarchi2011a}, the masses were derived in the same way, by comparing the $T_{\rm eff}$ and $L_{\rm bol}$ of the stars to MS isochrones from the same set of models, after reddening correction. {(Note that older PMS candidates and MS stars share generally the same position in the colour--magnitude diagram (CMD), the only difference being the presence or absence of H$\alpha$ excess emission with EW $>20$\,\AA.)} The resulting masses are in the range $\sim1.2-1.3$\,$M_\odot$, comparable with those of the PMS candidates. 

\section{Spectral extraction and flux calibration}

\begin{figure*}
\begin{center}
\resizebox{\hsize}{!}{\includegraphics{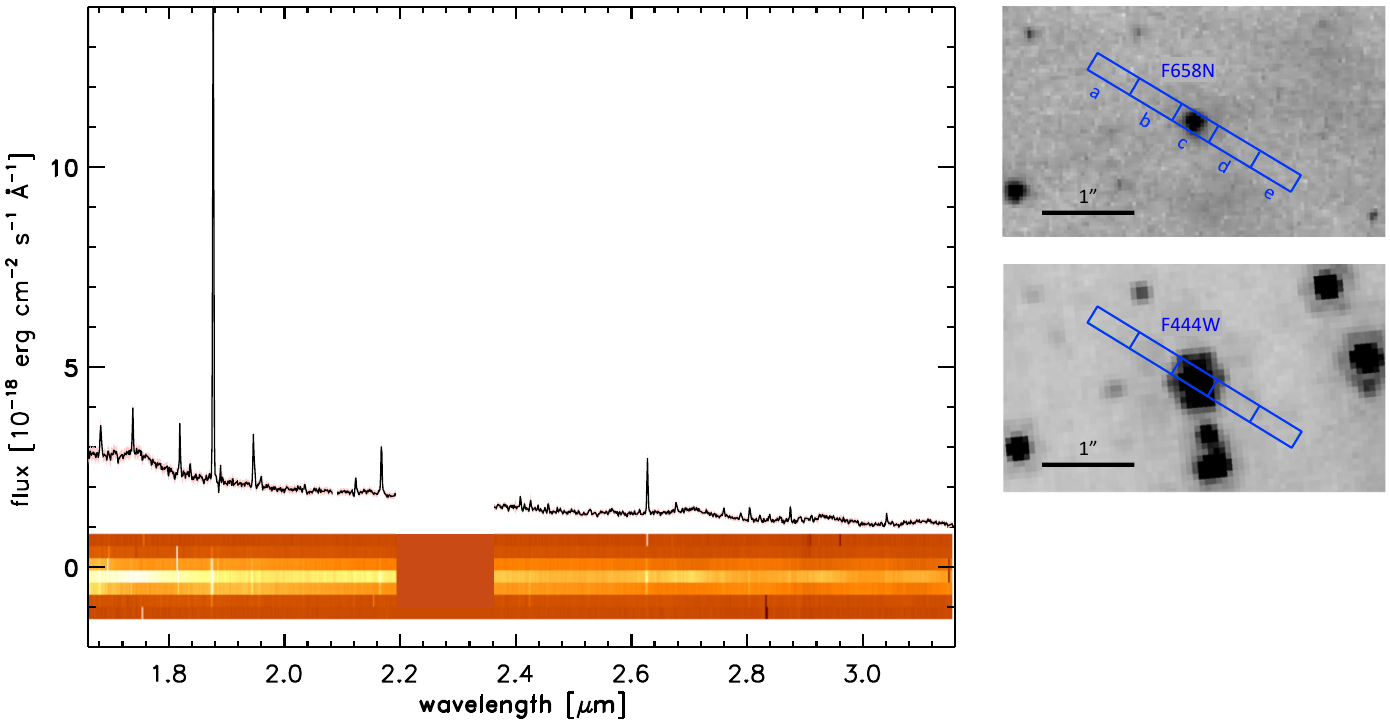}}
\resizebox{\hsize}{!}{\includegraphics{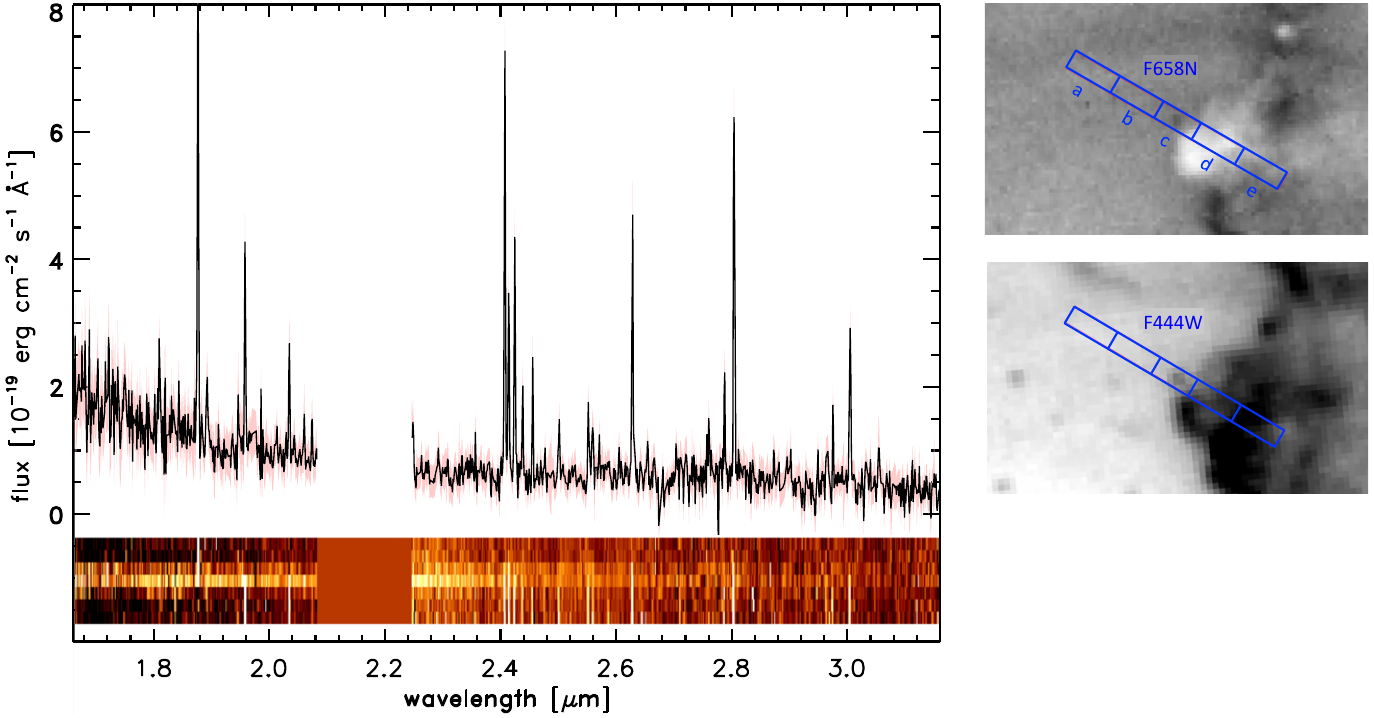}}
\caption{Spectra of sources 9884 (top panel) and 20408 (bottom panel). For each star we display the 1D spectrum (after background subtraction) and associated $1\,\sigma$ uncertainties (which are derived from standard error propagation through the reduction pipeline). In the bottom panel we show the 2D-rectified trace. The panels on the right-hand side show the {HST} F658N ($H\alpha$) and the NIRCam F444W filter negative images, with the three nodding positions of the NIRSpec slitlet array aperture shown in blue. There are three nodding positions, {namely ``abc", ``bcd", and ``cde", with the footprints of the five individual} microshutters labelled a to e {in the figure. Note that isolated bright pixels, such as those appearing in the 2D-rectified trace in the top panel near $1.8\,\mu$m and $2.6\,\mu$m, are the result of flags in the pipeline processing to exclude spurious pixels (e.g. hot pixels, cosmic-ray hits, saturation, no linearity correction, etc.). Once flagged, the pixels are excluded from the analysis and do not have any effect on the extracted one-dimensional spectra (same panel in Figure 1) that are used in this work.}}
\label{extfig2}
\end{center}
\end{figure*}

The exposures were processed with the NIRSpec ramp-to-slope pipeline, developed by the ESA NIRSpec Science Operations Team (SOT), which performs the following basic data reduction steps: bias subtraction, reference pixel subtraction, linearity correction, dark subtraction, and finally count-rate estimation, including jump detection and cosmic-ray rejection \citep{birkmann2022}. From the count-rate images the flux- and wavelength-calibrated spectra were obtained using the Stage 2 of the SOT NIRSpec Instrument Pipeline Software (NIPS; \citealt{nips2018}) to perform the following default operations \citep{alves2018}: subtract background, extract subimage containing the spectral trace, assign wavelength and spatial coordinates to each pixel therein, flat-fielding, and finally flux calibration. This procedure generates a rectified spectrum resampled on a regular two-dimensional (2D) grid, which in the case of our filter and grating combination (F170LP/G235M) produces images that span $7 \times 1414$ pixel$^2$, covering the wavelength range $1.66 - 3.16$\,$\mu$m, and $0\farcs46$ in the cross-dispersion direction. The 2D-rectified spectra of two of our sources are shown in Figure\,\ref{extfig2}. 

To optimise the SNR, we computed the one-dimensional (1D) spectrum of each star from the 2D-rectified product by coadding three pixel rows around the nominal source position, as given by the MSA planning tool. To account for any slit losses and to perform absolute calibration, we rescaled each spectrum in order to match the corresponding magnitude of each source in the F200W band as observed with NIRCam \citep{habel2024}. {The scale factor was derived by folding each observed spectrum through the NIRCam F200W band using the} {\em stsynphot} tool \citep{stsci2020} {and comparing the resulting synthetic magnitude with the one actually observed}. For stars 9685 and 40571 the magnitude in the F277W band was used instead, since these objects were not observed with the F200W filter. The NIRCam magnitudes are listed in Table\,\ref{tab1}.

Given the highly variable nebulosity in the region, background subtraction is a critical step. {We take as a reference the two strong HeI lines at $1.87\,\mu$m and $2.06\,\mu$m which are fully nebular in nature (the mass and $T_{\rm eff}$ of our targets are such that these lines are not expected to originate in their photospheres). We measure the flux of the HeI lines by integrating over their width, after subtraction of the continuum. This is done for the spectrum containing source+background and for the two background spectra sampled by the microshutters just above and below the source. We take as a template of the nebular background the spectrum from the microshutter where the integrated HeI line flux is closest to that in the spectrum with the source. We then rescale the background template (by multiplication) until the integrated HeI line flux agrees with that in the spectrum of the source, and finally subtract the rescaled template from the spectrum of the source. The multiplicative factor is always smaller than $1.2$ (i.e. 20\%) and typically ranges between $1.0$ and $1.1$, indicating small intensity variations in the nebular background around our sources, which our procedure corrects. }

In Figure\,\ref{extfig2}, the examples of sources 9884 and 20408 illustrate the two cases of uniform and highly varying backgrounds.  The outline of the three microshutters in the slitlet is shown on top of images in different bands, namely, $H\alpha$ (F658N) from the {HST} Advanced Camera for Surveys observations \citep{sabbi2007} and F444W from {JWST} NIRCam observations \citep{jones2023,habel2024}. Since there are three nodding positions, in total five microshutters are shown in projection.  

{Our final goal is to correct the H recombination lines in our sources, i.e. to remove the signatures of the nebular background. Therefore, the uncertainty introduced by this procedure depends on how well the observed HeI intensity variations match the variations in the H recombination lines. To assess this, we compare to one another the spectra of the two background microshutters (those above and below the source, namely, ``b'' and ``d'' in Figure\,\ref{extfig2}), after having registered them to have the same HeI line flux. For the three prominent H recombination lines in our study, intensity variations range from 1\,\% to 10\,\%, with typical difference of 5\,\% for Pa$\alpha$, 6\,\% for Br$\gamma$, and 5\,\% for Br$\beta$. For some targets, the HeI line flux in the background spectra matches that in the source spectrum (within Poisson uncertainties), so no rescaling of the background spectrum is needed before subtraction. This is the case when the nebular background is rather constant around the source, like for source 9884, where the background level in the $H\,\alpha$ band is rather uniform in the four regions probed by the nodding pattern around the source. The situation is very different for source 20408, where simple background subtraction without rescaling would not be appropriate.}

{We finally note that, by design, our nodding strategy places the spectrum of the source and the background spectra at the same location on the detector and in the field of view, thereby avoiding uncertainties caused by possible flat-fielding and throughput mismatches as well as nonresponsive pixels. }

After background subtraction, the spectra extracted from the three nodding positions of each source were median averaged to improve the SNR, and pixels flagged as saturated, nonlinear, or affected by high detector noise were excluded. 

\section{Spectroscopic diagnostics and physical properties of the stars}

\begin{figure*}
\centering
% GOOD!! \resizebox{\hsize}{!}{\includegraphics[angle=0,origin=b,height=13cm]{n346fig301e.pdf}}
\resizebox{\hsize}{7cm}{\includegraphics[angle=0,origin=b,height=13cm]{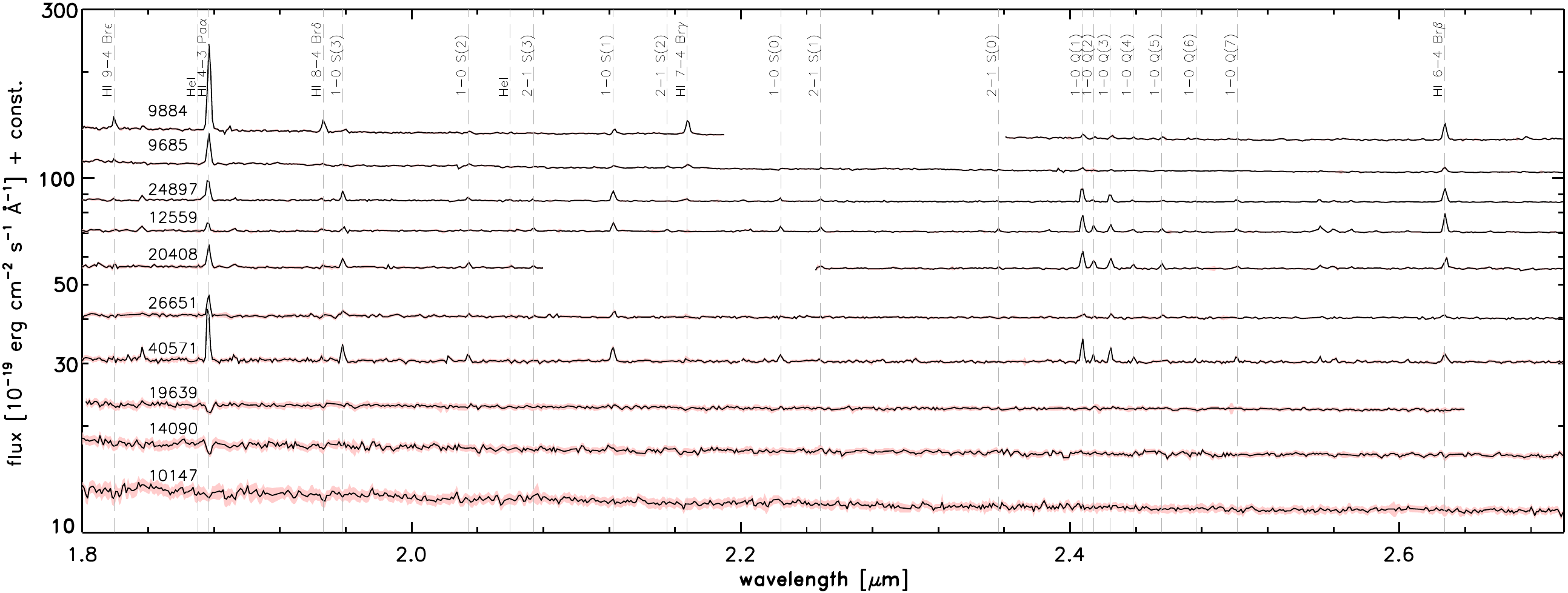}}
\caption{\small {Calibrated, background-subtracted spectra of seven PMS and three MS representative stars.} Object numbers correspond to Table\,\ref{tab1}. To improve visibility, the spectra are shifted vertically by offsets of $(11.5,10,8.5,7,5.5,4,3,2.2,1.6,1.1) \times 10^{-18}$ erg\,\,cm$^{-2}$\,s$^{-1}$\,\AA$^{-1}$. The vertical scale in this figure is compressed and the {H} absorption lines in the spectra of MS stars, although present, might be difficult to see. For a clearer view see Figure\,\ref{fig6}.}
\label{fig5}
\end{figure*}

The background-subtracted and combined spectra are shown in Figure\,\ref{fig5}, while the derived physical parameters of the stars are listed in Table\,\ref{tab1}. {The typical statistical uncertainty on the extracted spectra, after all pipeline steps and background subtraction, is $\sim 3\times10^{-20}$\,erg\,cm$^{-2}$\,s$^{-1}$\,\AA$^{-1}$. This is shown by the shaded areas around each spectrum in Figure\,\ref{fig5}. The corresponding SNR of the spectra (as defined by \citealt{stoehr2008}) ranges from 3 for source 40571 to 42 for source 9884. The typical uncertainty (median value) and range (from the 17$^{\rm th}$ to 83$^{\rm rd}$ percentile) of the measured line luminosities and EW are 2\,\% ($2-5\,\%$) for Pa\,$\alpha$, 6\,\% ($4-12\,\%$) for Br\,$\beta$, and 9\,\% ($2-16$\,\%) for Br\,$\gamma$ and are dominated by the uncertainty on the level of the underlying continuum. The absolute calibration of the spectra is based on NIRCam photometry in the F200W or F277W bands, which for our sources in the range $20<m_{200}<24$ has a typical uncertainty of $0.1$\,mag or $10$\,\% \citep{habel2024}.}

All PMS candidates show prominent H recombination lines in their spectra, in particular Pa\,$\alpha$ and Br\,$\beta$ (Br\,$\gamma$ is also in emission). For some of the PMS candidates, Br\,$\delta$ and Br\,$\epsilon$ are also detected. Conversely, the spectra of MS stars show H absorption lines but no emission lines. No He\,I nebular lines are present in any of these spectra, confirming good nebular background subtraction (see Section\,3). In Figure\,\ref{fig6}, we compare the spectrum of a PMS candidate of intermediate age (source 12559, $\sim 12$\,Myr, black line) with a representative MS spectrum (red line) obtained as the mean of the three MS stars in our sample. Three H absorption lines are clearly present in the MS spectrum (Br\,$\epsilon$, Pa\,$\alpha$, Br\,$\delta$), in contrast with the prominent emission features of the PMS star. 

In this section we {discuss the physical properties derived from the spectra, starting from those of MS stars. We also discuss and quantify three types of spectral features that are characteristic of young stellar objects, namely, atomic H recombination lines, molecular H excitation lines, and continuum near-infrared excess.}

\subsection{Metallicity, effective temperature, and surface gravity}

\begin{figure}
\centering
%\resizebox{0.5 \hsize}{!}{\includegraphics[angle=0,origin=b]{n346fig310.pdf}}
\resizebox{\hsize}{!}{\includegraphics[angle=0,origin=b]{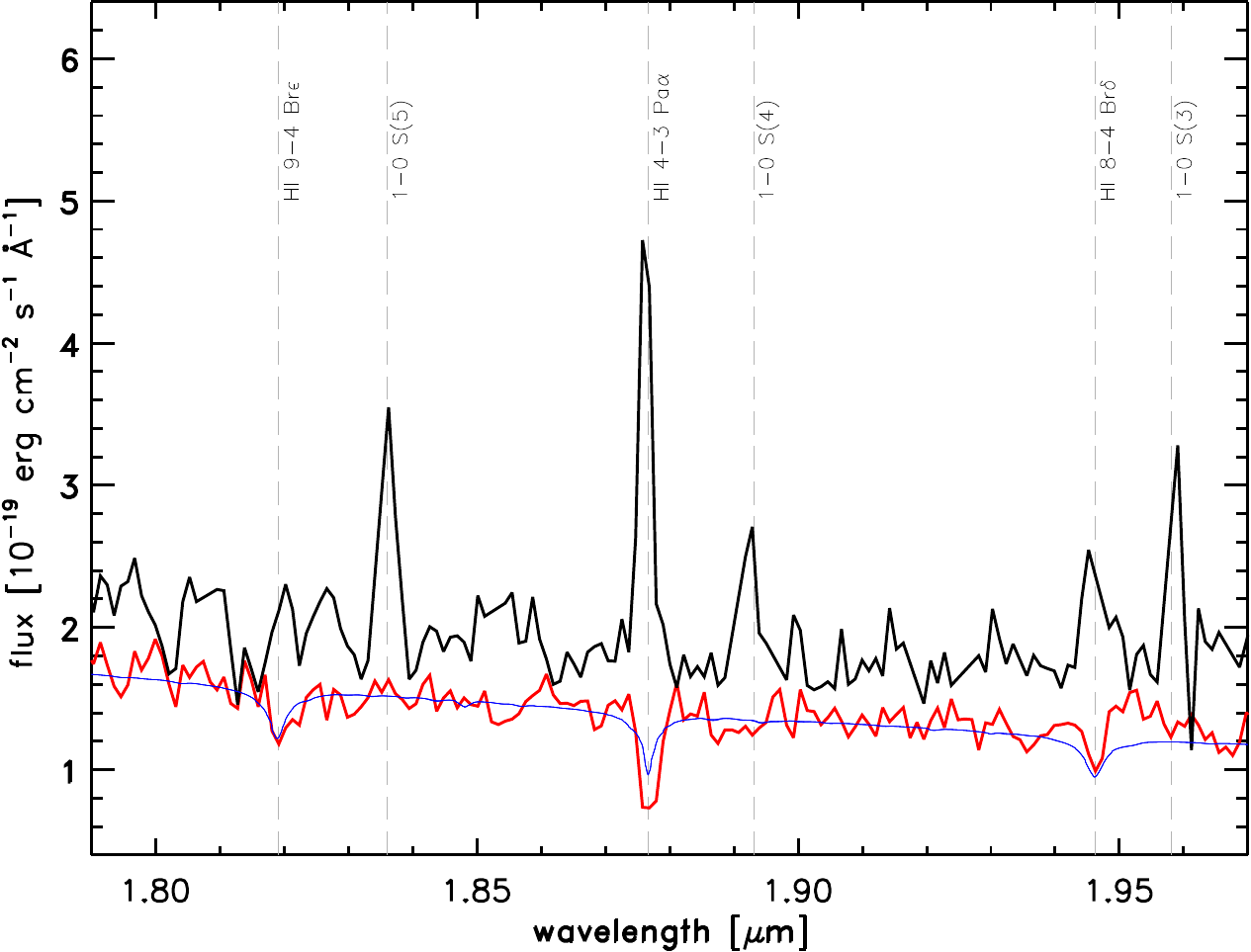}}
\caption{\small {Comparison of PMS and MS spectra.} The spectrum of PMS star 12559 (black line) is compared with the mean of the three MS spectra in our sample (red line), and with a model atmosphere with physical parameters appropriate for the MS objects (blue line).}
\label{fig6}
\end{figure}

No absorption spectral features from elements other than H are seen in any of our spectra. In fact, they are not expected because of the low metallicity of NGC\,346 and because of our limited resolving power ($R\sim1,000$). {For illustration purposes, the blue line in Figure\,\ref{fig6} shows, as an example, the theoretical MS spectrum \citep{hauschildt1999} for a stellar photosphere with metallicity $[M/H]=-0.5$ and $T_{\rm eff}=7,000$\,K, appropriate for these stars, and agrees well with the observations: no absorption lines other than those of H are present in the theoretical spectrum.}

{However, even without metal lines, we can set constraints on the effective temperature, surface gravity, and metallicity of the MS stars from the H absorption lines. To this end, we compared the spectra of sources 10147 and 14090 with the \citet{castelli2003} model atmospheres (suitably convolved and resampled to match the spectral resolution of the G235M grating of NIRSpec) because they are most appropriate for our observations in the relevant range of $T_{\rm eff}$ probed by these stars. We also tested the MARCS models \citep{gustafsson2008} and obtained typical cross-model uncertainties of $0.5$\,dex in $[Fe/H]$, 200\,K in $T_{\rm eff}$, and $0.2$\,dex in $\log g$. We assumed a value of 2\,km\,s$^{-1}$ for microturbulence, which is appropriate for the $T_{\rm eff}$ and $\log g$ values of our targets. As for the line list, we used as a reference the Vienna Atomic Line Database \citep{kupka2011}  in the range $1.7-3.2\,\mu$m. Both the MOOG (\citealt{sneden2012}; version 2019) and Spectroscopy Made Easy (\citealt{valenti1996}; version 2020) radiative transfer codes were tested, and with both codes we searched for the best fit in two ways: one in which the full wavelength range $1.71-1.97\,\mu$m is fitted, and one in which only the observed H lines are considered (which, unlike metal lines, are always clearly detected). In general, the MOOG code provides better fits. } 

{For star 11047, when considering only H absorption lines in the stated range, the best fit is obtained for $T_{\rm eff}=8,200\pm500$\,K, $\log g=4.5\pm0.5$, and $-1.0 < [Fe/H]< -0.5$. When instead the fit is extended to the whole wavelength range $1.71-1.97\,\mu$m, the preferred values are $T_{\rm eff}=7,700\pm300$\,K, $\log g=4.4\pm0.5$, and $-1.0 < [Fe/H]< -0.5$. In both cases $v \sin i$ cannot be constrained because our $\sim 300$\,km s$^{-1}$ resolution is too coarse. In both cases, solar metallicity can be excluded, but the lack of any metal lines does not allow us to set firmer constraints on the actual metallicity, other than to say that the range $-1.0 < [Fe/H]< -0.5$ is consistent with the observations. However, we highlight here that the lack of metal lines applies to all sources in our sample, even the brightest ones for which the SNR in the continuum exceeds 30. This in itself is an indication that the metallicity is low.}

{In the case of star 14090, if only H absorption lines are considered, the parameter values providing the best fit are $T_{\rm eff}=7,700\pm300$\,K, $\log g=4.6\pm0.5$, and $-1.0 < [Fe/H]< -0.5$. If the complete $1.71-1.97\,\mu$m spectral range is considered, the values are $T_{\rm eff}=7,500\pm400$\,K, $\log g=4.3\pm0.5$, and $-1.0 < [Fe/H]< -0.5$. Also for this star can $v \sin i$ not be constrained.}

{We find that the values of $T_{\rm eff}$ and their uncertainties are fully consistent with those derived by \cite{demarchi2011a} from the photometry of these stars in the $V$ and $I$ bands, as listed in Table\,\ref{tab1}. Furthermore, comparison of the corresponding model atmospheres \citep{castelli2003} allows us to confirm also the values of $A_V$ derived by \cite{demarchi2011a} from the photometry (see Table\,\ref{tab1} and Section\,5). While we cannot use the same approach to set constraints on $T_{\rm eff}$ and the other stellar parameters for PMS stars, because the H lines are in emission, the excellent match for MS stars gives us confidence about the photometrically-derived $T_{\rm eff}$ also for the PMS sources in our sample.}

\begin{table*}[ht]
\centering
\footnotesize
\caption{{Properties of the stars in the Lupus and Taurus sample}. The columns correspond to: source identification (ID), class, effective temperature ($T_\mathrm{eff}$), age, stellar mass ($M_*$), and EW of the Br\,$\gamma$, Br\,$\zeta$, and Pa\,$\beta$ lines ($1\,\sigma$ uncertainties are shown in parenthesis). All spectral lines have positive EW, indicating absorption. }

\tabcolsep=0.05cm
\begin{tabular}{lcccccccc}
\hline
ID &  Class &  $T_{\rm eff}$ &  Age & Mass &  $W_{\rm Br\gamma}$ & $W_{\rm Br\zeta}$ & $W_{\rm Br\eta}$ &  $W_{\rm Pa \beta}$ \\
   &        &  $K$           &  Myr & $M_\odot$ & \AA\, & \AA\, & \AA\, & \AA\, \\
\hline
RXJ1556.1-3655 &  II & 3800 &  8 & $0.6$ &$0.12$ ($0.03$)&                & &               \\
Sz122          & III & 3500 &  7 & $0.4$ &$0.94$ ($0.14$)&                & & \\
HQ Tau         &  II & 5000 & 10 & $2.0$ &                &$0.55$ ($0.06$)&$0.08$ ($0.05$)&$0.58$ ($0.08$) \\
V1115 Tau      & III & 4500 & 10 & $1.0$ &                &$0.24$ ($0.06$)&$0.14$ ($0.05$)&$0.25$ ($0.02$)\\
\hline
\label{exttab2}
\end{tabular}
\end{table*}

\subsection{H recombination lines and mass accretion}

The prominent H emission lines in Figure\,\ref{fig5} demonstrate for the first time that the NGC\,346 stars with $H\alpha$ excess emission in { HST} photometry are indeed actively accreting PMS objects. Following the relationships for a sample of 91 PMS objects in Lupus \citep{alcala2017}, we converted the Br\,$\gamma$ line luminosities $L(Br\gamma)$ of the PMS spectra, corrected for extinction, into the accretion luminosities $L_{\rm acc}(Br\gamma)$ {(for more details on the procedure see Appendix\,A and \citealt{rogers2024})}. In Table\,\ref{tab1} these can be compared with the $L_{\rm acc}(H\alpha)$ accretion luminosities obtained by applying the relationships of \cite{alcala2017} to the $L(H\alpha)$ line luminosities measured photometrically \citep{demarchi2011a} for the same objects. {The uncertainty on the derived $L_{\rm acc}$ is about $0.25$\,dex (see \citealt{alcala2017}).}

Except for stars 26651 and 12559, where the weak Br\,$\gamma$ line implies an uncertain luminosity, the two independent sets of $L_{\rm acc}(H\alpha)$ and $L_{\rm acc}(Br\gamma)$ values, based on measurements taken 18 years apart, agree to better than a factor of 2, consistent with the accretion rate variability observed for PMS objects \citep{gahm2008,herczeg2009,biazzo2012,venuti2014}. 

Using the available \citep{demarchi2011a} stellar masses and ages (Table\,\ref{tab1}), we can derive mass accretion rates (see Appendix\,A for the details). For stars older than $\sim10$\,Myr, the typical accretion rate is $\sim10^{-8}\,M_\odot$\,yr$^{-1}$, which is in good agreement with that estimated for similar objects in the Magellanic Clouds \citep{demarchi2010,demarchi2011a,demarchi2013,demarchi2017,biazzo2019,carini2022,tsilia2023,vlasblom2023}, and yet about an order of magnitude larger than for {{stars of similar mass and age}} in nearby Galactic star-forming regions \citep{hartmann2016}. 

%No empirical relationships between $L_{\rm acc}$ and $L(Pa\alpha)$ or $L(Br\beta)$ exist in the literature. 

A preliminary comparison of the observed line ratios to those predicted by temperature- and electron-density-dependent models \citep{storey1995} for case B hydrogen recombination \citep{baker1938} suggests $T_{\rm eff}\simeq5,000$\,K and $N_e\simeq10^{10}$\,cm$^{-3}$. These values are consistent with those predicted by magnetospheric accretion models for classical T Tauri stars \citep{martin1996,muzerolle1998}, although the temperature is slightly lower than the $6,000-12,000$\,K range predicted by \cite{muzerolle2001}. 

\subsubsection{Comparison with older Galactic PMS stars}

We have compared the IR spectra of the older PMS sources in NGC 346 with those of $\sim 10$ Myr old stars in two Milky Way star-forming regions, Taurus and Lupus. We selected two sources in Taurus (HQ Tau and V1115 Tau) and two in Lupus (RXJ1556.1-3655 and Sz122) with ages in that range. The spectra of the sources in Taurus come from TNG/GIARPS NIR observations \citep{alcala2021} and the corresponding physical parameters were derived by \cite{gangi2022}. The NIR spectra of the sources in Lupus are taken from VLT/XSHOOTER observations \citep{alcala2014,alcala2017} and the physical parameters of the sources are discussed in \cite{frasca2017}. The PMS nature of these Taurus and Lupus sources was assessed by those authors on the basis of the NIR excess and the ages were determined by comparing the stellar photometry with theoretical PMS isochrones, which is the same approach followed for the NGC 346 sources (\citealt{demarchi2011a} and Section\,5).

Table\,\ref{exttab2} summarises the physical parameters of the four Milky Way sources (as published in the papers mentioned above), to which we have added the EW of some representative H recombination lines in the NIR, as measured by us on the calibrated background-subtracted spectra provided in those papers. 

\begin{figure*}
\centering
\resizebox{\hsize}{!}{\includegraphics[angle=0,origin=b]{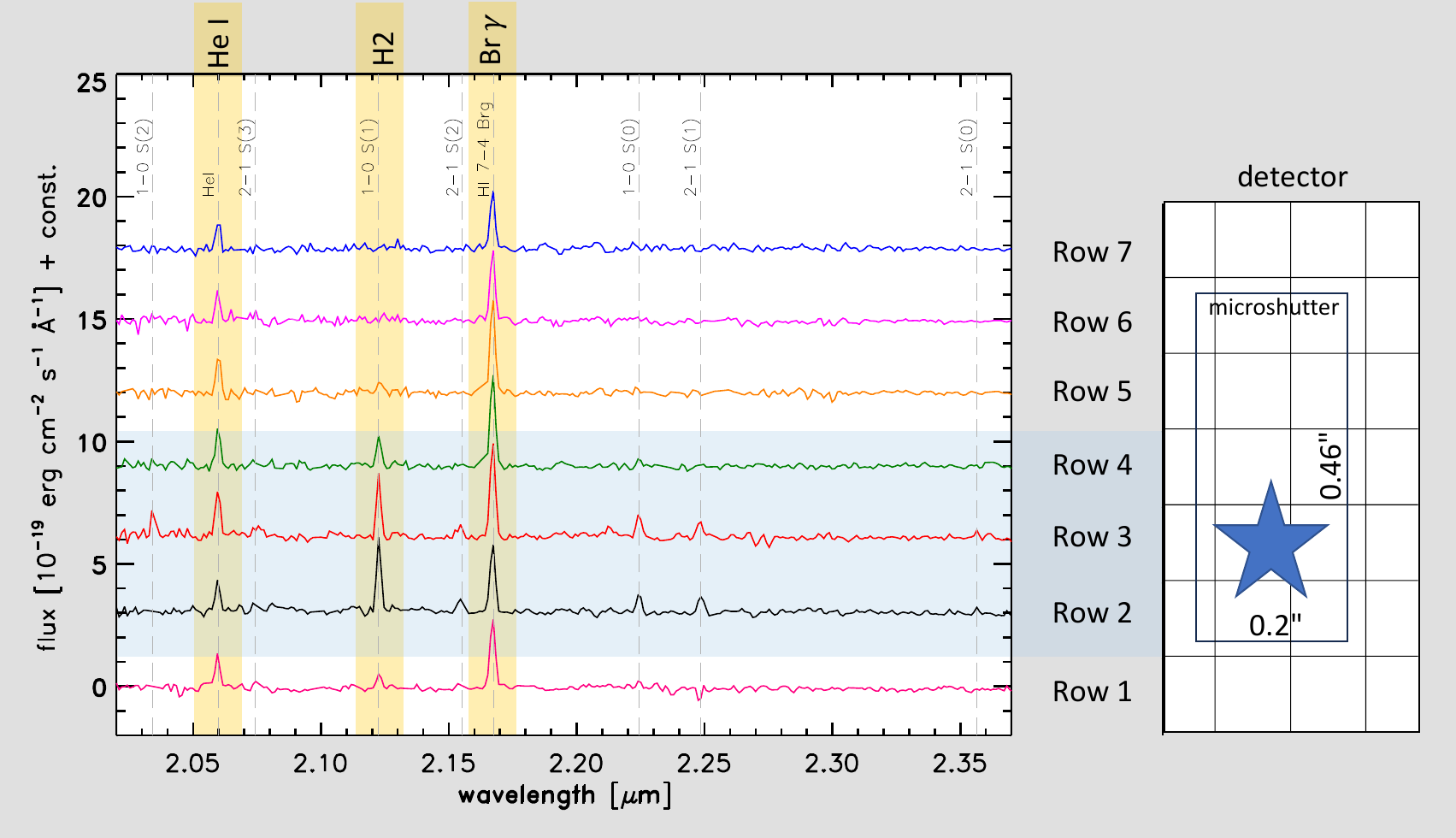}} % {aasjan23slide15.pdf}}
\caption{\small {Localised H$_2$ excitation lines around source 40571.} Left: source spectra along the seven individual pixel rows of the rectified trace, without background subtraction. Right: projected position of the source in the microshutter. The $v=1\rightarrow0~$S$(1)$ excitation line is stronger at the specific location of the source, while nebular lines are present across the full slit length.}
\label{fig7}
\end{figure*}

The XSHOOTER spectra of the sources in Lupus sample the Br$\gamma$ line. For both objects the line is found to be in absorption, with EW of $0.12$\,\AA\, (with 1-$\sigma$ uncertainty of $0.03$\,\AA) for RXJ1556.1-3655, and $0.94$\,\AA\, ($0.12$\,\AA) for Sz122. The GIARPS spectra of the Taurus sources do not sample Br$\gamma$, so we looked at nearby H recombination lines, starting from Br\,$\zeta$ (Br--10 at $1.736$\,$\mu$m, which is also covered in the NIRSpec spectra but is too faint for our sources, except for 9884 and 9685). This line is also seen in absorption, with EW of $0.55$\,\AA, and $0.24$\,\AA, for HQ Tau and V1115 Tau, respectively. Since telluric lines are very pronounced around Br\,$\zeta$, and some residual absorption might remain even after subtraction, we also inspected the Br\,$\eta$ (Br--11 at $1.681\,\mu$m) and Pa\,$\beta$ lines (not included in the NIRSpec spectra of the NGC 346 sources). The Br\,$\eta$ and Pa\,$\beta$ lines are also in this case clearly seen in absorption, with EW of $0.08$\,\AA\ and $0.55$\,\AA\ for HQ Tau, and $0.14$\,\AA\ and $0.25$\,\AA\ for V1115 Tau, respectively. As a result, we expect that also Br$\gamma$ and Pa$\alpha$ will be in absorption for these sources. 

On the other hand, objects with comparable ages in NGC\,346 (ID 24897, 12559, and 20408) show strong emission in the Br$\gamma$ and Pa$\alpha$ lines, with EW of order 10\,\AA\, for Br$\gamma$ and hundreds of \AA\, for Pa\,$\alpha$, clearly indicating that the stars are still accreting material from the circumstellar environment. In summary, while PMS stars with ages of about 10 Myr show strong signs of accretion in NGC\,346, no such signs are detected in stars of similar age in small nearby Galactic star-forming regions with solar metallicity.

\subsection{H$_2$ excitation lines and outflows}

The background-subtracted spectra of the PMS stars in Figure\,\ref{fig5} reveal a conspicuous number of molecular hydrogen (H$_2$) excitation lines. In particular, the lines corresponding to the rovibrational transitions $v=1\rightarrow0$\,S$(0)$, $1\rightarrow0$\,S$(1)$, $1\rightarrow0$\,S$(2)$, $1\rightarrow0$\,S$(3)$, $2\rightarrow1$\,S$(1)$, $2\rightarrow1$\,S$(2)$, $2\rightarrow1$\,S$(3)$, as well as $1\rightarrow0$\,Q$(1)$ through to $1\rightarrow0$\,Q$(7)$ are observed. Since the spectra are background subtracted, the molecular gas must be spatially associated with the stars. 

By knowing the exact location of the stars in the microshutter, we can set constraints on the distribution of the H$_2$ molecular material.  Figure\,\ref{fig7} shows the spectra of source 40571, without background subtraction; the side panel indicates the position of the source inside the microshutter (as provided by the MSA configuration file), which is intermediate between pixel rows 2 and 3 in the extracted rectified trace (note that rows 1 and 7 of the trace are only partly illuminated). While the He\,I and Br\,$\gamma$ nebular emission lines extend across the full height of the slit, the spatial extent of the $v=1\rightarrow0$\,S$(1)$ emission line is more limited around the specific location of the source. The same is true for the other six PMS candidates, indicating that H$_2$ is likely associated with circumstellar material.

{The extent of the H$_2$ emission in the spatial direction is about $0\farcs1$ (1 pixel) and is not resolved. Therefore, we can only set an upper limit to the extent of the H$_2$ circumstellar structure (possibly an outflow in the disc or a filament), which at the distance of NGC\,346 corresponds to $\sim6,000$\,AU. This is clearly just an upper limit, and we cannot set any constraints on the shape or structure at the origin of the H$_2$ emission, but we note that this size is within a few times the typical $\lesssim1,000$\,AU size of discs around Galactic PMS stars measured in Taurus \citep{najita2018}. } 

Some of these H$_2$ lines are observable from the ground in nearby protostars and their intensity ratios are used to characterise the physical conditions and excitation mechanism in their circumstellar environments \citep{greene2010}. Here, we provide a preliminary characterisation of the environments and excitation mechanisms from the intensity ratios of the  S$(1)$ lines in the $v=2\rightarrow1$ and $v=1\rightarrow0$ transitions.  For star 12559, the intensity ratio is $0.47$ suggesting nonthermal ultraviolet electron pumping, for which \cite{gredel1995} predict a ratio of $0.54$ (the value is higher if extinction is present). The intensity ratio of the S$(3)$ lines in the $v=2\rightarrow1$ and $v=1\rightarrow0$ transitions agree with this scenario. For stars 9685, 24897, and 40571 the ratios are in the range $0.11-0.18$ characteristic of LTE molecular gas at temperatures of $\sim2,200-2,700$\,K, as commonly seen in shocks from outflows. Stars 26651, 9984, and 20408 were analysed for the $2\rightarrow1$ S(3) line ratio because the S(1) line was affected by bad pixels or not covered in the wavelength range of the spectra.  These three stars also have line ratios consistent with shock excitation of $H_2$ gas in LTE at $\sim2,200-2,700$\,K. 

In summary, the intensity ratios of the S$(1)$ and S$(3)$ lines in the $v=2\rightarrow1$ and $v=1\rightarrow0$ transitions suggest that the likely excitation mechanism for one of the stars (12559) is ultraviolet electron pumping, while for the other objects shocks appear to be the cause. The measured $H_2$ line ratios in these stars point to excitation mechanisms that likely trace a variation in underlying physical processes from outflows and circumstellar disc material commonly seen in young stars in the Milky Way. %A detailed analysis of the $H_2$ excitation lines for the PMS candidates in Figure\,\ref{fig5} will be presented in a forthcoming paper.

\subsection{Near-infrared excess and circumstellar dust}

\begin{figure}
\centering
%\resizebox{0.5 \hsize}{!}
\resizebox{\hsize}{!}
{\includegraphics[angle=0,origin=b]{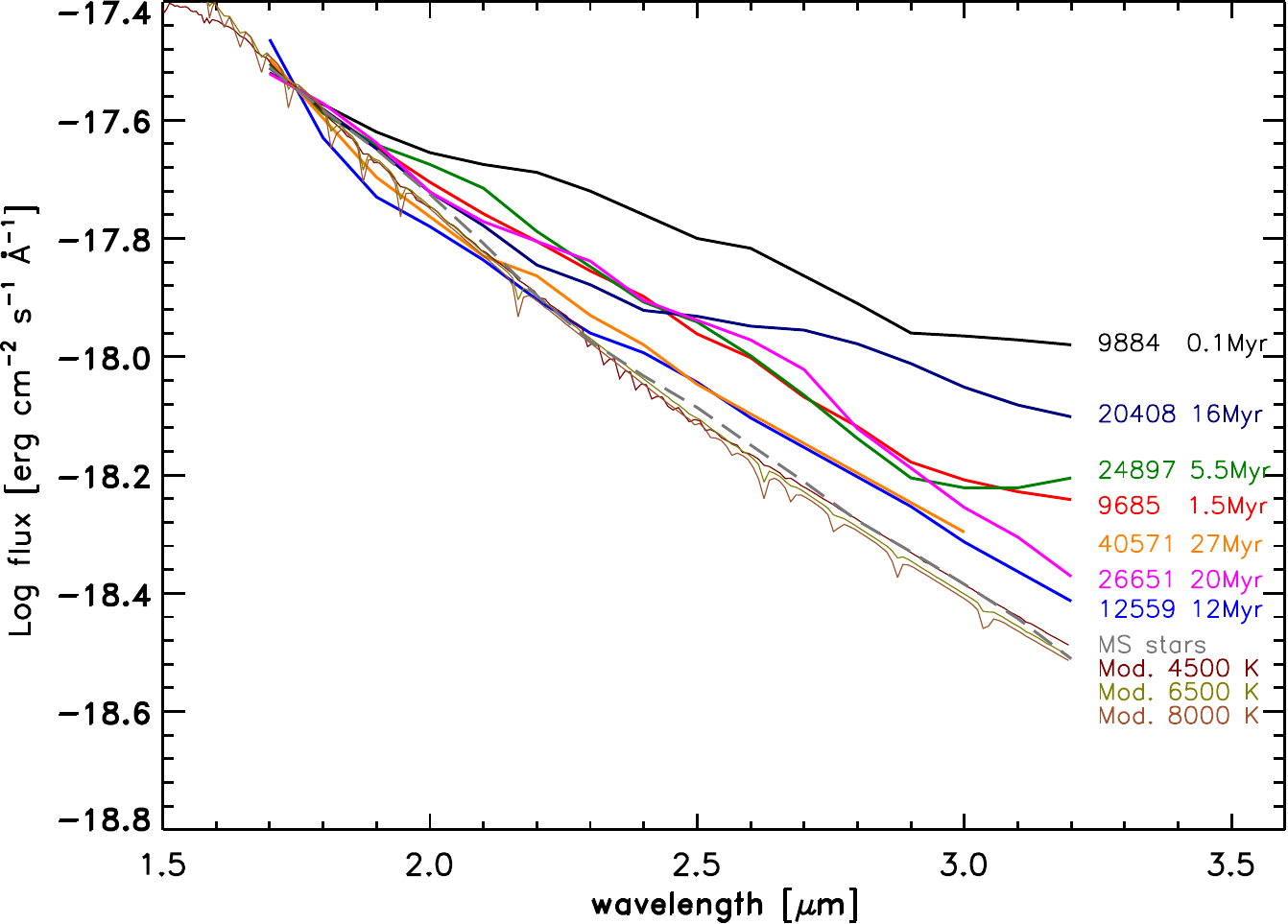}}
\caption{\small {SEDs (fitted continua) of the dereddened sources, registered at $1.75\,\mu$m.} The spectra of PMS stars are shown individually (thick colour lines as per the legend), while the three MS stars are averaged together (grey dashed line), as in Figure\,\ref{fig6}. The thin lines correspond to theoretical model spectra for metallicity $[M/H]=-0.5$ and $T_{\rm eff}=4,500$\,K, $6,500$\,K and $8,000$\,K, respectively. }
\label{fig8}
\end{figure}

Finally, the spectral energy distributions (SEDs) of the PMS stars in our sample present NIR excess, caused by dust emission, which provides additional evidence of the presence of circumstellar discs. In Figure\,\ref{fig8} we compare the SEDs of the stars obtained by fitting the continuum of the spectra in Figure\,\ref{fig5} using a linear interpolation with a step of $\sim 0.1\,\mu$m. The derived SEDs were corrected for reddening, using the $A_V$ values in Table\,\ref{tab1} with the extinction law of \cite{rieke1985}, and are normalised at $1.75\,\mu$m for comparison. The grey dashed line corresponds to the average MS star spectrum shown in Figure\,\ref{fig8} and matches closely the model spectra \citep{bessell1998} for metallicity $[M/H]=-0.5$, surface gravity $\log g=4.5$ and $T_{\rm eff}$ of 4,500\,K, 6,500\,K and 8,000\,K (see the legend). This comparison confirms that in this wavelength and temperature range the spectral shape in the Planck tail of the distribution is not particularly sensitive to the exact $T_{\rm eff}$ value. The differences observed in the SEDs of the PMS stars, with respect to those of the MS objects,  clearly indicate the presence of NIR excess. This further confirms that the stars with excess emission in the H recombination lines are not standard MS stars. If their continuum matched that of MS stars, the emission lines might simply be the result of an incomplete subtraction of the nebular background. Instead, the markedly different continuum shape provides independent evidence that these objects are not MS stars and are still surrounded by circumstellar material.

{All PMS candidates except for source 12559 reveal NIR excess also in the NIRCam photometric bands F115W, F200W, and F277W  \citep{habel2024}, with excess values ranging from $0.25$ to $1.3$\,mag compared to the colours of the MS stars in our sample. In general terms (see Table\,\ref{tab1}), younger PMS candidates have typically larger NIR excess, with some scatter. The scatter, however, is not surprising: although the mass of the circumstellar disc decreases over time, the amount of NIR excess does not directly scale with age because of the effects introduced by different disc geometry and morphology, including inclination \citep{garufi2018}.} For instance, the presence of spirals and shadows is associated with a high NIR excess, while the presence of rings in the discs appears to cause low NIR excess \citep{garufi2018}, irrespective of the actual PMS age. Therefore, a pronounced NIR excess for an older PMS candidate, like star 20408, can be expected.

\section{Discussion: A matter of age}

\begin{figure*}
\centering
%\resizebox{0.5 \hsize}{!}
\resizebox{\hsize}{!}
{\includegraphics[angle=0,origin=b,height=12cm]{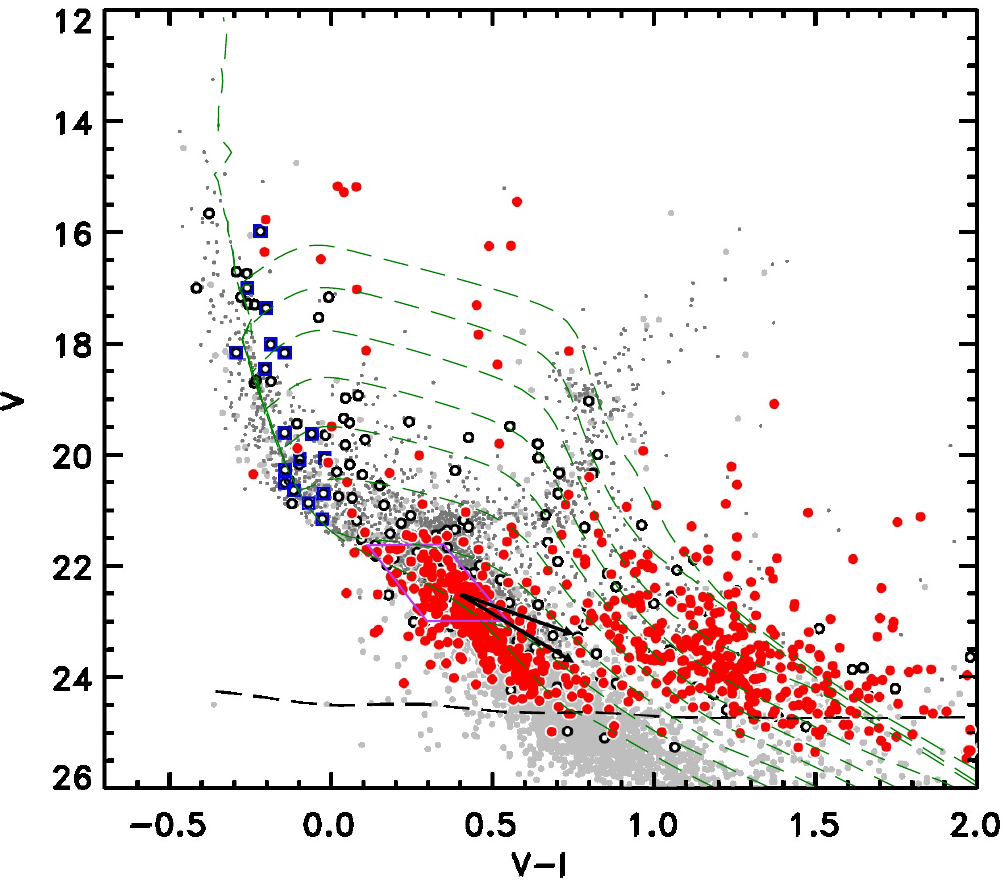}
 \includegraphics[angle=0,origin=b,height=12cm]{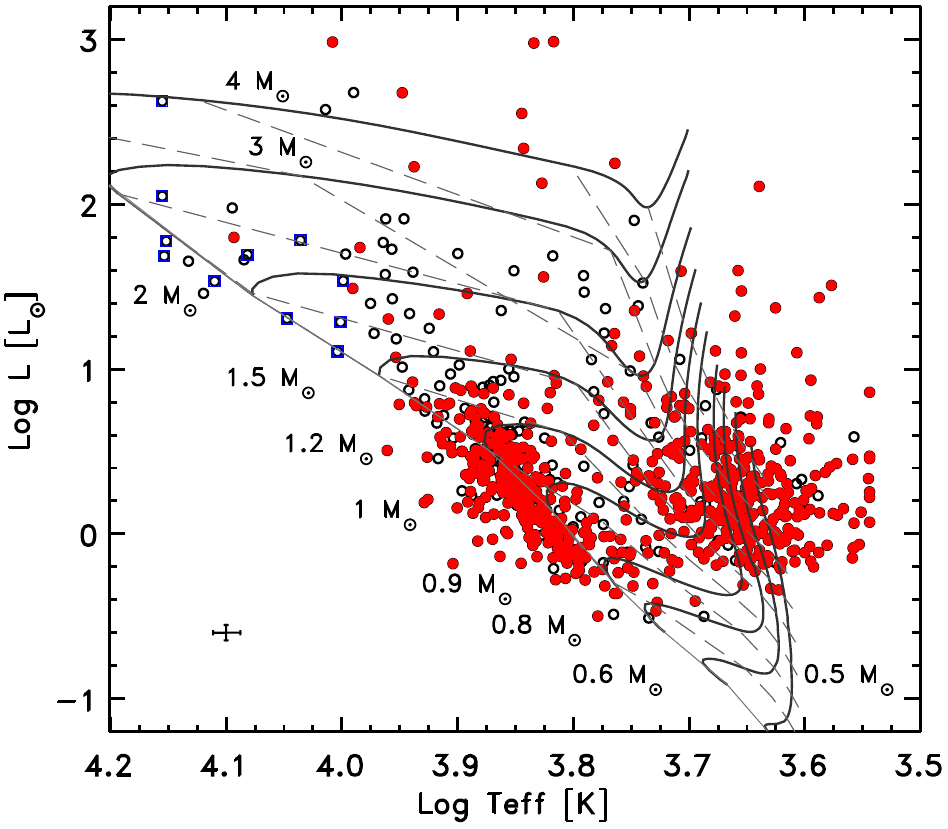}}
\caption{\small Left: CMD of the stars in NGC\,346 \citep{demarchi2011a}. All magnitudes are corrected for extinction. Objects with excess emission in H$\alpha$ at the $4\,\sigma$ level or higher are shown as open circles. A total of 694 of them (red dots) also have H$\alpha$ excess emission with EW in excess of 20\,\AA\ or more than 50\,\AA\ for stars hotter than $10,000$\,K. Squares correspond to objects with H$\alpha$  excess emission and EW of less than 50\,\AA\ that are potential Be stars. The thick black dashed line corresponds to the 50\,\% completeness limit of the photometry in the $V$ and $I$ bands \citep{sabbi2007}. The majority of other stars in the field (grey dots) do not show H$\alpha$ excess emission. {Thin green dashed lines correspond to isochrones from \cite{tang2014} for $Z=0.002$ and ages of $0.125$, $0.25$, $0.5$, 1, 2, 4, 8, 16, and 32 Myr from right to left. The arrows show representative reddening vectors for colour excess $E(V-I)=0.5$ and two values of the total-to-selective extinction ratio: $R=3.1$ (standard Galactic extinction law, {top arrow}), and $R=4.5$ as measured in the low-metallicity massive starburst cluster 30 Dor {(bottom arrow)}}. Right panel: HRD of the stars with H$\alpha$ excess at the $4\,\sigma$ level or higher. Thick solid lines show the evolutionary tracks from the Pisa group \citep{deglinnocenti2008,tognelli2011} for metallicity $Z = 0.002$ and masses from $0.5$ to $4$\,\Msolar, as indicated. {The corresponding isochrones are shown as thin dashed lines, for the same ages as in the left panel.} Note that the constant logarithmic age step has been selected such that the typical distance between isochrones is larger than the uncertainties (indicated by the typical error bars in the bottom-left corner of the HRD). The purple polygon is discussed in Section\,5.4. }
\label{fig9}
\end{figure*}

In this section we discuss the ages of the PMS stars in our sample, how they were derived from the photometry, and their uncertainties. Some elements of this discussion were presented by \cite{demarchi2011a}, but here we address the matter in more detail and through a number of independent arguments, all of which point to an old age for a {significant} fraction of the accreting PMS stars in NGC\,346.

{A search for PMS stars in NGC\,346 \citep{demarchi2011a} revealed approximately ${\sim}690$ candidates with strong H$\alpha$ excess emission in their photometry, corresponding to H$\alpha$ EW $>20$\,\AA. In the $V, V-I$ CMD, after reddening correction, these PMS candidates form two distinct groups, a red group and a blue group, with a median $V-I$ colour separation of $0.75$\,mag at all magnitudes. When the $T_{\rm eff}$ and $L_{\rm bol}$ of these stars, derived from the photometry, are compared in the Hertzsprung--Russell diagram (HRD) with widely-used PMS models \citep{siess2000,tognelli2011,tang2014}, the $0.75$\,mag colour separation translates into a clearly bimodal age distribution, with about half of the objects younger than ${\sim} 2$\,Myr and the other half older than ${\sim} 8$\,Myr, including over 200 stars with ages in the range $15-30$\,Myr. To help readers appreciate the systematically different locations of the younger and older PMS candidates in the CMD and HRD, we have included in the left and right panels of Figure\,\ref{fig9}, respectively, the CMD and the HRD as published by \cite{demarchi2011a}, with those authors' permission.} These figures are relevant because they were used to select the spectroscopic targets for the investigation reported here. 

{Hereafter, we will refer to these two groups of PMS stars as ``older PMS candidates'' and ``younger PMS candidates'', respectively. The former include the stars with $H\alpha$ excess emission in Figure~\ref{fig9} with colour $0.1 \la V-I \la 0.8$, which \cite{demarchi2011a} showed to be older than ${\sim} 8$\,Myr (median age 20\,Myr), to have masses in the range $0.7 - 1.5$\,\Msolar, and to be located primarily in the outskirts of the association. The latter include objects with $H\alpha$ excess emission in Figure~\ref{fig9} with colour $0.8 \la V-I \la 2.0$, younger than ${\sim} 2$\,Myr (median age $0.9$\,Myr), with masses in the range $0.4 - 4.0$\,\Msolar, and located near the centre of the association, in compact sub-clusters.}

{In the following subsections we address} three specific points, which are however intertwined, namely, the location of the PMS sources in the CMD, their reddening, and their spatial distribution. We also explore how representative the older emitting PMS sources are of the older PMS population in NGC\,346 as a whole.

\subsection{CMD and photometric uncertainties}

{Here, we discuss the location of the PMS sources in the CMD and show that their small photometric uncertainties, combined with the low metallicity of the NGC\,346 cluster, allow us to set stringent constraints on the relative ages of the younger and older PMS stars.}  

{The typical photometric uncertainty in the $V$ and $I$ bands for the PMS stars in the catalogue of \cite{sabbi2007} used by \cite{demarchi2011a} is $0.02$\,mag, which translates into typical uncertainties of about $1-2$\,\% on $T_{\rm eff}$ and of $\sim 6$\,\% on $L_{\rm bol}$. In that work, the value of $T_{\rm eff}$ was derived by comparing the observed $V-I$ colour, after correction for reddening, with those stemming from the model atmosphere of \cite{bessell1998} for metallicity $[M/H]=-1.0$ and gravity $\log g=4.5$, as appropriate for the PMS sources in NGC\,346, and for the specific { HST} bands used in those observations (see \citealt{demarchi2010,demarchi2011a} for details). The stellar parameters derived spectroscopically in Section\,4 attest to the accuracy of the effective temperature obtained from the photometry. Bolometric luminosities are based on the adopted SMC distance modulus of $18.92 \pm 0.03$, corresponding to ${\sim} 61$\,kpc \citep{hilditch2005,keller2006}. While it has long been known that the SMC extends {about 5 kpc} along the line of sight (e.g. \citealt{gardiner1991,nidever2011,yanchulova2017,murray2024}), there is no evidence that this is the case for the specific NGC\,346 star forming region. The diameter of the region is about 60\,pc, and one expects the extent along the line of sight to be similar. This will have no measurable effects on the dispersion in the photometry and hence on the derived value of $L_{\rm bol}$.}

\begin{figure}
\centering
\resizebox{\hsize}{!}
{\includegraphics[angle=0,origin=b]{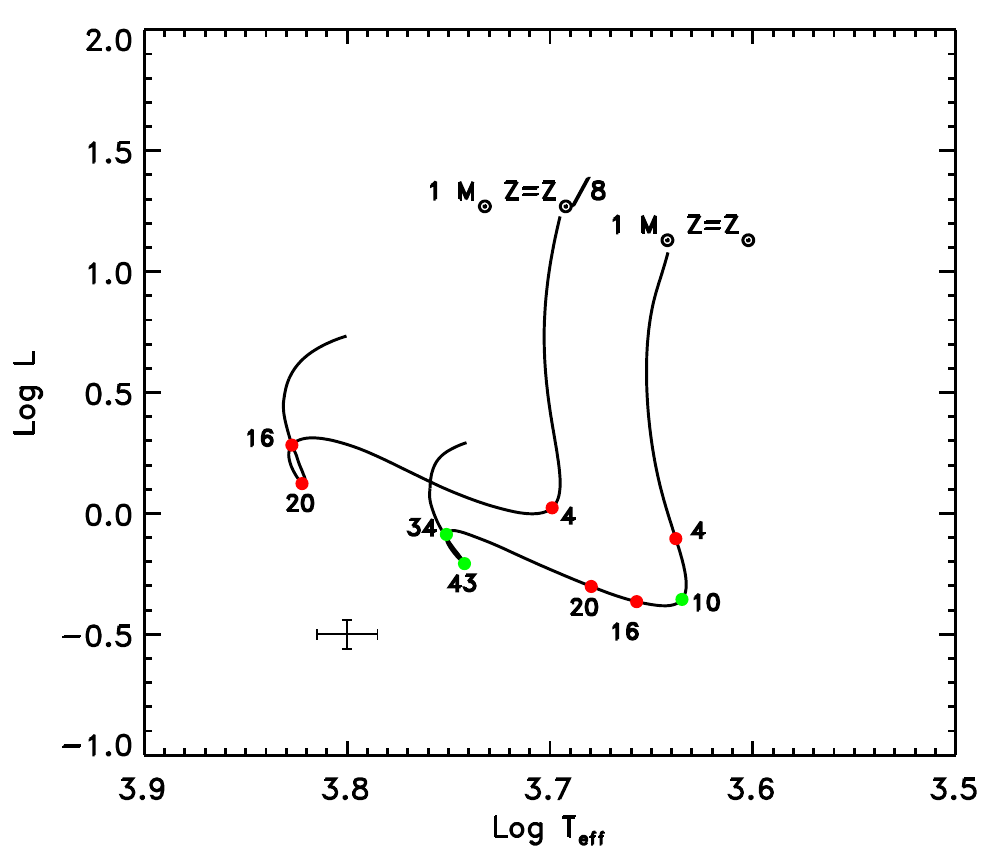}}
\caption{\small PMS evolutionary tracks from \cite{tognelli2011} for a 1\,\Msolar\, star and two different metallicities, as indicated. Numbers along the tracks correspond to PMS ages in Myr. At $Z=1/8\,Z_\odot$, the transition from the bottom of the Hayashi track to the MS occurs earlier and more rapidly than at $Z_\odot$. For reference, we also show the error bars from the right panel of Figure\,\ref{fig9}.}
\label{fig10}
\end{figure}

{Figure\,\ref{fig9} shows that older PMS candidates with H$\alpha$ excess emission (which we here show to have excess emission also in the Paschen and Brackett lines) appear very close to the location of the MS, shown by the solid grey line in the right panel of Figure\,\ref{fig9}.} Objects of this type are not observed in nearby Galactic star-forming regions. However, their position in the HRD is fully consistent with theoretical models of stellar evolution in low-metallicity environments. At the metallicity of NGC\,346 ($Z=1/8\, Z_\odot$ or $Z\simeq0.002$), theoretical evolutionary tracks \citep{deglinnocenti2008,tognelli2011} show that a 1\,\Msolar\,star reaches the MS after about 20\,Myr, as we illustrate in Figure\,\ref{fig10}. The time is about 27 Myr for a $0.9$\,\Msolar\,object like source 40571 in Table\,\ref{tab1}. At solar metallicity, models by the same authors show that this time is twice as long, or about 43\,Myr for a 1\,\Msolar\,star. {We verified that both the MIST \citep{dotter2016} and PARSEC \citep{bressan2012} evolutionary tracks would produce very similar results.}

More importantly, even if the lifetime of circumstellar discs did not depend on metallicity, Figure\,\ref{fig10} implies that in low-metallicity environments there is a higher probability of finding {PMS stars close to the MS which are still actively accreting. Indeed, photometric investigations of star-forming regions across the Magellanic Clouds have systematically shown this result over the past decade} \citep{demarchi2010, demarchi2011a, demarchi2011b, demarchi2013, demarchi2017, biazzo2019, carini2022, tsilia2023, vlasblom2023}. Obviously, if disc lifetimes increase at lower metallicity, as our observations appear to indicate (see Section\,6), the likelihood of finding still accreting PMS stars close to the MS is even higher. 

{Figure\,\ref{fig10} {presents an additional important implication for uncertainties on ages, which are derived through isochrone comparison and as such are affected by uncertainties on} $T_{\rm eff}$ and $L_{\rm bol}$. The latter in turn depend on uncertainties in the photometry and on the extinction correction (typical uncertainties on $\log T_{\rm eff}$ and $\log L_{\rm bol}$ are shown by the error bars in Figure\,\ref{fig10}). None of these uncertainties depend on metallicity. Since at lower metallicity stars transit more quickly across the HRD, a given uncertainty on $\log T_{\rm eff}$ and $\log L_{\rm bol}$ will translate into a lower uncertainty on the age. We quantify this in Figure\,\ref{fig10}, where PMS evolutionary tracks for 1\,\Msolar\,stars and two metallicities are compared. At $Z=1/8\,Z_\odot$, the star leaves the bottom of the Hayashi track at $t=4$\,Myr and reaches the MS region 12\,Myr later, having become about 1500\,K warmer, at an average rate of 150\,K\,Myr$^{-1}$. At solar metallicity, the {time spent on the radiative part of the PMS track is} 24\,Myr and $T_{\rm eff}$ increases by about 1300\,K, so the average transfer rate is about 55\,K\,Myr$^{-1}$. During the transfer phase, which accounts for the majority of the PMS lifetime, age uncertainties are about three times smaller at $Z=1/8\,Z_\odot$ than at solar metallicity {(the actual factors, for a 1\,\Msolar{} star, are $2.7$, $2.5$, and $2.6$ for the PISA, MIST, and PARSEC tracks, respectively)}. This analysis shows that PMS stars in NGC\,346 form two distinct age groups.}

\subsection{Reddening}

{One might suppose that the two groups of PMS candidate stars in the CMD and HRD of Figure\,\ref{fig9} belong in fact to the same population (i.e., have the same age) but simply appear in different regions of the CMD because they {are affected by} different reddening values, which have not been properly corrected. In the following, we show why this is not the case.}

\text{First of all, the} extinction in this field is very limited, as the shape of the Red Clump indicates (see CMD in Figure\,\ref{fig9}). This was shown by \cite{demarchi2011a} who inspected the photometry of ${\sim} 600$ massive stars in the cluster and derived the $E(V-I)$ colour excess for each of them {and by \cite{habel2024}, who found tightly constrained RCs in near- and mid-infrared CMDs with JWST photometry}. Since PMS candidates and massive stars have a similar spatial distribution in this field, \cite{demarchi2011a} derived a statistical reddening correction for each PMS candidate from the median value of the extinction towards the five nearest massive star neighbours. That study showed that, even though some differential reddening is present in the field, with $E(V-I)$ ranging from $0.1$ and $0.25$\,mag (17 and 83 percentiles, respectively), the $V-I$ colour separation between red and blue PMS stars in the CMD of Figure\,\ref{fig9} is five times wider than the median colour excess in the field, $E(V-I)\simeq0.16$. {These values are in turn consistent with those reported by \cite{hennekemper2008}, who found $A_V$ in the range $0.2 - 0.4$ mag in the densest central regions.} 

The arrows in the CMD show the reddening vectors for two extinction laws: the one for the standard diffuse Galactic interstellar medium (e.g. \citealt{cardelli1989,fitzpatrick1999}), with $R=3.1$ corresponding to $A_V=1.7 \times E(V-I)$, {is shown by the upper arrow}; {the lower arrow corresponds to the extinction law} measured in the massive starburst cluster 30 Dor in the Large Magellanic Cloud \citep{demarchi2014,maizapellaniz2014,demarchi2016,fahrion2023}, with $R=4.5$ corresponding to $A_V=3.0 \times E(V-I)$. {The latter is likely to be more typical of starburst regions, in particular at low metallicity. } It is clear that reddening cannot displace PMS stars from one group to the other within the observed magnitude range, since in the CMD the reddening vectors run parallel to the separation between the two groups {(note that to improve visibility the vectors are shown for $E(V-I)=0.5$ or three times the median value in the field).} To bridge the two groups, an unusually shallow reddening vector would need to be invoked, $A_V=1.1 \times E(V-I)$. Furthermore, extinction can only shift objects to redder colours and fainter magnitudes in the CMD; therefore the bluer stars can only have less extinction than the red ones, not more. So the older PMS ages derived for the bluer stars cannot be attributed to insufficient reddening correction. {In other words, if the reddening correction were insufficient, the ages would be underestimated, {but less so for a steep reddening vector.} }

\subsection{Spatial distribution of younger and older PMS stars}

{Also the spatial location of the PMS stars in the field suggests an intrinsic physical difference between the two groups of PMS candidates in Figure\,\ref{fig9}. In theory, episodic accretion in the early stages of evolution could make a small fraction of PMS stars appear bluer in the CMD, thereby mimicking an older age \citep{baraffe2009,baraffe2010,vorobyov2013,baraffe2017,kuffmeier2018}. However, \cite{demarchi2011b} showed that the spatial distribution of the redder PMS stars is different from that of the bluer PMS objects: the former are more centrally concentrated and coincide with the small subclusters identified by \cite{sabbi2007}, while the latter are more widely distributed and not associated with the subclusters. Similar results were found using \textit{JWST} photometry \citep{jones2023}.} Recently \citet{sabbi2022} demonstrated that the two populations also exhibit different kinematic patterns. 

{Therefore, in NGC\,346 we can exclude episodic accretion as the main cause of the two separate groups of PMS objects in Figure\,\ref{fig9}, owing to the spatial segregation of the two groups in the field. Indeed, it is difficult to see how this effect, which must apply with equal probability to all stars regardless of their position in space, could explain the correlation observed between the colours of PMS stars and their spatial location in the cloud. We conclude that the two groups of PMS stars are not members of the same population and their clearly distinct positions in the CMD can only be explained by a difference in age.}

\subsection{Accreting and nonaccreting older PMS stars}

A final question is whether the older PMS sources with strong hydrogen emission are typical of the older PMS population of NGC\,346 as a whole, or whether they represent  the tail of a larger distribution of coeval stars that are no longer accreting, such as, for instance, in the case of the ${\sim} 15$\,Myr old TW Hya \citep{herczeg2014}. To address this question, we turn to the sample of 330 NGC\,346 PMS stars older than 8\,Myr and with $H\alpha$ EW larger than 20\,\AA\, as identified photometrically by \cite{demarchi2011a}.  {This stringent selection inevitably leaves out some similarly aged sources that are weakly accreting or no longer accreting.}   

Since the area observed with { HST} around NGC\,346 \citep{sabbi2007,demarchi2011a} is dominated by SMC field stars, in order to limit statistical fluctuations we focus our analysis on the region where the density of older PMS stars is higher. We defined an ellipsoidal region near the centre of NGC\,346 in such a way as to contain 50\,\% of the older PMS candidates. 
The region spans about 3 square arcmin. {We further restrict the sample by considering stars with $V<23$ since this ensures completeness better than 50\,\% in the $V$, $I$, and $H\alpha$ bands. A total of 63 PMS stars satisfy this additional condition and in the $V, V-I$ CMD they occupy an area containing an additional 483 sources with no H$\alpha$ excess emission (the area is outlined by the purple polygon in the left panel of Figure\,\ref{fig9}). } 

{If all the 483 sources were nonaccreting PMS members of NGC346, this would imply a $\sim 12\,\%$ fraction of accreting PMS objects at ages of $10-30$ Myr. Given that many of these objects are likely to be SMC field stars, not related to the NGC\,346 cluster, this is a lower limit and suggests that the older accreting PMS stars are significantly more than a tail in the distribution.}

{A proper determination of the actual fraction of accreting PMS stars requires a detailed statistical analysis of the field star contamination, including a thorough assessment of the photometric completeness and extinction, both of which are known to vary across the field \citep{sabbi2007,hennekemper2008}. The study is beyond the scope of this paper and we plan to undertake it in a future work.}

\section{Conclusions}

{Thanks to} the excellent combination of sensitivity and spatial resolution of NIRSpec on JWST {we have been able} to observe spectrscopically a sample of extragalactic solar-mass candidate PMS stars, {for the first time}. {The candidates were identified through H$\alpha$ excess emission in their photometry}. {Our} new data set confirms that these objects are indeed PMS stars, with the spectral signatures of active accretion {and the presence} of colder molecular gas and of dust in circumstellar material. The spectroscopic confirmation of the PMS nature of the photometrically selected candidates suggests that it will be possible to extend the use of narrowband imaging methods to reliably identify and study star-forming objects in a variety of environments, also in galaxies further away in the Local Group. 

The spectral features presented in Section\,4 are common to {both younger and older} PMS stars in our sample, with ages spanning from $\sim 0.1$ to $\sim 30$\,Myr. {Independently} of the age spread, all our objects reveal typical PMS signatures: prominent H recombination lines implying large accretion luminosities, H$_2$ excitation lines suggesting molecular material in the immediate vicinity of the stars, and NIR excess witnessing the presence of dusty circumstellar discs. {This suggests} that in NGC\,346 circumstellar discs live longer than typical discs in the Milky Way and PMS stars can accrete at sustained rates for over 20\,Myr. {This is in contrast with what is observed in the Milky Way, where at solar metallicity, discs around stars of $\sim$ 1\,\Msolar\ dissipate very quickly \citep{pfalzner2022}.} 

Our results {could} appear at odds with the outcome of a study of two small low-metallicity ($\sim0.1\,Z_\odot$) star-forming regions in the extreme outer Galaxy \citep{yasui2009,yasui2010}, {where {a} small fraction of sources with near-IR excess suggested very short disc lifetimes, even shorter than in the solar neighbourhood.} However, subsequent studies of other low-metallicity Galactic clusters indicate that the {actual fraction of disc-bearing stars in the outer Galaxy} is controversial \citep{yasui2016,cusano2011} and subject to large age uncertainties \citep{guarcello2021}. Nor is it yet clear what is the exact role of stellar feedback on the evolution and lifetime of circumstellar discs. \cite{ramirez2023} studied circumstellar discs in the massive Galactic cluster NGC\,6357: while they were expecting the discs to be more affected by the UV radiation from massive stars than in the much smaller Lupus star-forming region, the authors found that the spectra in the two regions are quite similar. 

{Thus}, from an observational point of view, there is currently no compelling reason to believe that at low metallicity circumstellar discs are short-lived. Furthermore, the prevailing theories requiring  short lifetimes appear to be problematic for the formation of the planets known to exist around low-metallicity stars \citep{sigurdsson2003, niedzielski2009,setiawan2010}, and might question the validity of the core-accretion model for the formation of planetesimals \citep{johnson2012}. For planets to form, dust must settle in the midplane before the disc dissipates, but the settling timescale is longer at the low dust-to-gas ratios expected in these environments and it increases with distance from the star. If disc dissipation in low-metallicity stars were as quick as initially reported \citep{yasui2009}, only very small rocky planets close to the star could form \citep{johnson2012}. 

{In conclusion}, our results {indicate} that, at the low metallicities typical of the early Universe, the disc lifetimes {may be longer than what is observed} in nearby star-forming regions, thus allowing more time for giant planets to form and grow than in higher-metallicity environments. {This may have significant implications for our understanding of the formation of planetary systems in environments similar to those in place at Cosmic Noon.}

\begin{acknowledgments}
{We are very grateful to an anonymous referee for providing insightful and stimulating comments that have helped us improve the presentation of our results.} This work is based on observations made with the NASA/ESA/CSA James Webb Space Telescope. The data were obtained from the Mikulski Archive for Space Telescopes at the Space Telescope Science Institute, which is operated by the Association of Universities for Research in Astronomy, Inc., under NASA contract NAS 5-03127 for \textit{JWST}. These observations are associated with program JWST-GTO-1227. The specific observations analyzed in this work can be accessed via doi: 10.17909/tatn-1114. MM and NH acknowledge support through a NASA/JWST grant 80NSSC22K0025, and MM, NH, and LL acknowledge support from the NSF through grant 2054178. ON acknowledges support from STScI Director’s Discretionary Fund. KF acknowledges support through the ESA research fellowship. KB acknowledges support through Mini Grant INAF 2022 TRAME@JWST (TRacing the Accretion Metallicity rElationship with NIRSpec@JWST). OCJ acknowledges support from an STFC Webb fellowship. CN acknowledges the support of a STFC studentship (2645535). MM, NH, and LL acknowledge that a portion of their research was carried out at the Jet Propulsion Laboratory, California Institute of Technology, under a contract with the National Aeronautics and Space Administration (80NM0018D0004). ASH is supported in part by an STScI Postdoctoral Fellowship. We thank Brunella Nisini, Teresa Giannini, Juan Alcal\'a, and Manuele Gangi for discussions. 
\end{acknowledgments}


%\vspace{5mm}
\facilities{JWST(NIRSpec, NIRCam)}


\software{eMPT  \citep{bonaventura2023}, NIPS  \citep{nips2018}}

%% Appendix material should be preceded with a single \Appendix command.


%\appendix
%\section{Accretion luminosity and mass accretion rate}

\vspace{0.5cm}
\centerline{APPENDIX}

\section*{A. Accretion luminosity and mass accretion rate}

We derived the accretion luminosities listed in Table\,\ref{tab1} following the relationships of \cite{alcala2017} between accretion luminosity and line luminosity, of the type $\log(L_{\rm acc}/L_\odot)=a\times\log(L_{\rm line}/L_\odot)+b$, where $a$ and $b$ are specific to each line and $L_{\rm line}$ is the line luminosity corrected for extinction. The values of $L({\rm Br}\gamma)$ are those measured in this work, while those of $L(H\alpha)$ are taken from \cite{demarchi2011a}. The coefficients (and their uncertainties) are $a=1.13\,(0.05)$ and $b=1.74\,(0.19)$ for H$\alpha$, and $a=1.19\,(0.10)$ and $b=4.02\,(0.51)$ for Br$\gamma$.  

The extinction correction at the wavelength of Br\,$\gamma$ was computed from the $A_V$ values measured by \cite{demarchi2011a} for each star (see Table\,\ref{tab1}) and converted to the $K$ band by adopting the standard relationship \citep{rieke1985,vandehulst1949} $A_K\simeq 0.1\,A_V$. The typical reddening is $A_K =0.05 \pm 0.01$ and after extinction correction $L(Br\gamma)$ becomes about 5\,\% brighter. Although the effect introduced by extinction is comparable to the typical calibration uncertainty, it is systematic and we corrected for it.

The mass accretion rates (Table\,\ref{tab1}) were computed from the freefall equation

\begin{equation}
L_{\rm acc} \simeq \frac{G\,M_*\,\dot M_{\rm acc}}{R_*} \left(1 -
\frac{R_*}{R_{\rm in}}\right)
\label{eq5}
\end{equation}

\noindent
where $G$ is the gravitational constant, $M_*$ and $R_*$ the mass and photospheric radius of the star, and $R_{\rm in}$ the inner radius of the accretion disc, from where the accretion flow is supposed to start. The values of $M_*$ and $R_*$ (listed in Table\,\ref{tab1}) are from \cite{demarchi2011a}. The value of $R_{\rm in}$ depends on how exactly the accretion disc is coupled with the magnetic field of the star and is therefore uncertain. Following \cite{gullbring1998}, we adopt $R_{\rm in} = 5\,R_*$ for all PMS objects. 

Concerning the uncertainties, besides those on the accretion luminosity also those on the $R_*/M_*$ ratio contribute to the uncertainty on $\dot M_{\rm acc}$. The uncertainty on $R_*/M_*$ is dominated by the reddening correction \citep{demarchi2010}. Given the typical $A_V=0.5$ extinction of our sources, adopting a conservative $0.1$\,mag uncertainty on $A_V$ results in an uncertainty of $\sim 3\,\%$ on $R_*/M_*$. This is smaller than the systematic uncertainly on the mass alone, which \cite{demarchi2010} estimate to be about 20\,\%, because the effects of the extinction on $R_*$ and $M_*$ are correlated. Therefore, the $0.25$\,dex uncertainty on the $L({\rm line}) - L_{\rm acc}$ conversion dominates the uncertainty on $\dot M_{\rm acc}$. The overall cumulative relative uncertainty on the mass accretion rate is estimated to be $\sim 0.42$\,dex \citep{alcala2017}.


\end{document}